\documentclass[aps,prl,showpacs,twocolumn]{revtex4}

\usepackage{amssymb}
\usepackage{epsfig}
\usepackage{graphicx}

\begin{document}

\title{Controlling soliton interactions in Bose-Einstein condensates
by synchronizing the Feshbach resonance and harmonic trap}
\author{Xiao-Fei Zhang$^{1,2}$, Qin Yang$^{1,3}$, Jie-Fang Zhang$^4$, X. Z. Chen$^5$, and W. M. Liu$^{1}$}
\address{$^1$Beijing National Laboratory for Condensed Matter Physics,
Institute of Physics, Chinese Academy of Sciences, Beijing 100080,
China}

\address{$^2$Department of Physics, Honghe University, Mengzi 661100, China}

\address{$^3$School of Electrical and Mechanical Engineering, Jiaxing University, Jiaxing 314001, China}

\address{$^4$Institute of Nonlinear Physics, Zhejiang Normal University, Jinhua 321004, China}
\address{$^5$Key Laboratory for Quantum Information and Measurements of
Ministry of Education, School of Electronics Engineering and
Computer Science, Peking University, Beijing 100871, China}
\date{\today}

\begin{abstract}
We present how to control interactions between solitons, either
bright or dark, in Bose-Einstein condensates by synchronizing
Feshbach resonance and harmonic trap. Our results show that as
long as the scattering length is to be modulated in time via a
changing magnetic field near the Feshbach resonance, and the
harmonic trapping frequencies are also modulated in time, exact
solutions of the one-dimensional nonlinear Schr\"{o}dinger
equation can be found in a general closed form, and interactions
between two solitons are modulated in detail in currently
experimental conditions. We also propose experimental protocols to
observe the phenomena such as fusion, fission, warp, oscillation,
elastic collision in future experiments.
\end{abstract}


\maketitle

\section{I. Introduction}

Observation of Bose-Einstein condensates (BECs) in gases of weakly
interacting alkali-metal atoms has stimulated intensive studies of the
nonlinear matter waves. One of the central questions in this field
is how to explore properties of BECs. It is known that interatomic
interactions greatly affect a number of properties of BECs,
including both static (such as the size, shape and stability) and
dynamic ones (like the collective excitation, soliton, and vortex
behavior, etc.). A common practice to change the interaction
strength, even its sign, is to modulate the $s$-wave scattering
length, $a_s$, by using the Feshbach resonance with a tunable
time-dependent magnetic field $B(t)$
\cite{Moerdijk,Roberts,Stenger,Inouye,Cornish,Donley,Regal,Volz03}:
$a_s(t)= a_{\infty} [1 -\triangle/(B(t)-B_0)]$, where $a_{\infty}$
is the off-resonance scattering length, $t$ is the time, $B_0$ and
$\triangle$ are the resonance position and width, respectively.
This offers a good opportunity for manipulation of atomic matter
waves and nonlinear excitations in BECs. In real experiments,
various forms of the time dependence of $B(t)$ have been explored
\cite{K. E. Strecker,C. A. Regal,M. Greiner}, observation of dark
and bright solitons have been reported. In theoretical studies,
several forms of time-varying scattering lengths have been
proposed and treated separately, such as the exponential function
$e^{\lambda t}$ \cite{F. Kh. Abdullaev, V. M. Perez-Garcia, Z. X.
Liang}, or the periodic function $g_0 +g_1 \sin(\Omega t)$
\cite{H. Saito, D. E. Pelinovsky, G. D. Montesinos, Konotop95, M.
Matuszewski}, and so on.

In the present paper, we will consider the general case with
arbitrary, time-varying scattering length $a_s(t)$, and discuss
how to control dynamics of solitons in BECs by synchronizing the
Feshbach resonance and harmonic trap in current experimental
conditions. We first obtain a family of exact solutions to the
general nonlinear Schr\"{o}dinger equation with an external
potential and arbitrary time-varying scattering length $a_s(t)$,
then further discuss how to control the interaction of solitons
including the bright and dark solitons. We observe several
interesting phenomena such as fusion, fission, warp, oscillation,
elastic collision in BECs with different kinds of scattering
length correspond to different real experimental cases.

\section{II. the model and soliton solutions}

Consider condensates in a harmonic trap $V({\bf r})= m
\omega^2_{\bot} (y^2+z^2)/2 +m\omega^2_1 x^2/2$, where $m$ is
atomic mass, $\omega_{\bot}$ and $\omega_1$ the transversal and
axial frequency, respectively. Such a trap can be realized, for
instance, as a dipole trap formed by a strong off-resonant laser
field. In the mean-field theory, the dynamics of BEC at low
temperature is governed by the so-called Gross-Pitaevskii (GP)
equation in three-dimensions. If $\omega_{\bot} \gg |\omega_1|$,
it is reasonable to reduce the GP equation for the condensate wave
function to the quasi-one-dimensional (quasi-1D) nonlinear
Schr\"{o}dinger equation \cite{K. D. Moll, G. Fibich, Garcia98,
Kevrekidis04, Brazhnyi04},
\begin{equation}
i\frac{\partial \psi}{\partial t}= -\frac{1}{2} \frac{\partial^2
\psi}{\partial x^2} +\frac{a_s(t)}{a_B} |\psi|^2 \psi
+\frac{\omega^2_1}{2 \omega^2_{\bot}} x^2 \psi, \label{GP}
\end{equation}
where the time $t$ and coordinate $x$ are measured, respectively,
in units of $\omega^{-1}_{\bot}$ and $a_\bot$, with $a_\bot \equiv
\sqrt{\hbar/m \omega_{\bot}}$; $\psi$ is measured in units of
$1/(\sqrt{2\pi a^2_\bot a_B})$, with $a_B$ as the Bohr radius. The
key observation of the present paper is that if we allow the axial
frequency of the harmonic trap becomes also time dependent
$\omega_1= \omega_1(t)$, and require it to satisfy the following
integrability relation with the scattering length $a_s(t)$, we
have
\begin{equation}
-\frac{1}{a_s(t)} \frac{d^2 a_s(t)}{d t^2} + \frac{2}{a_s^2(t)}
(\frac{d a_s(t)}{d t})^2 +\frac{\omega^2_1(t)}{\omega^2_{\bot}}
=0.  \label{IR}
\end{equation}

Then, the nonlinear GP equation (\ref{GP}) with time-varying
coefficients can be reduced to the standard nonlinear
Schr\"{o}dinger equation and exactly solved, with the following
general solution:
\begin{equation}
\psi(x,t)= exp[-\int \limits^{t}_{t_0} \Gamma(t)dt] \phi(X,T)
\exp[i\Gamma(t) x^2], \label{SO}
\end{equation}
where $\phi$ is an arbitrary function of $X$ and $T$, with new
spatial and temporal variables $X= A_1 \exp[-2 \int
\limits^{t}_{t_0} \Gamma(t) dt] x$, $T= \frac{A^2_1}{2} \int
\limits^{t}_{t_0} \exp[-4\int \limits^{t}_{t_0} \Gamma(t') dt']
dt$. $A_1$ is a real constant, which together with $\Gamma(t)$ are
determined by
\begin{equation}
 a_s(t)= \pm a_B A^2_1 \exp[-2\int \limits^{t}_{t_0} \Gamma(t)dt]. \label{A}
\end{equation}

Meanwhile, the trapping frequency can also be expressed in terms
of $\Gamma(t)$ given by
\begin{equation}
 \frac{\omega_1^2(t)}{\omega_{\bot}^{2}} = -2\Gamma_{t} - 4\Gamma(t)^{2}. \label{W}
\end{equation}

From Eqs. (\ref{A}) and (\ref{W}), we see that both the scattering
length and the trapping frequency can be expressed in terms of the
$\Gamma$ function. It shows that the trapping potential can become
repulsive, we will give a detailed exposition of this situation in
the following section. Since the trapping potential we consider
here is the cigar-shaped harmonic potential (here-after, the
frequency $\omega_{\bot}$ is not varying), once the function of
scattering length is determined, the $\Gamma$ function and the
function of the trap potential can also be determined. Note that
the exact solution $\psi(x,t)$ can be obtained for arbitrary time
dependence of $a_s(t)$, since we can always choose an appropriate
time-dependent axial frequency, $\omega_1(t)$, to satisfy the
integrability relation.

When the interatomic interaction is attractive, i.e., $a_s(t)<0$,
Eq. (\ref{GP}) has bright $N$-soliton solutions. The simplest case
for studying soliton interactions is the two bright solitons
solution, for which $\psi(x,t)$ is expressed by Eq. (\ref{SO})
with $\phi(X,T)$ given by
\begin{equation}
\begin{array}{ll}
\phi(X,T)= 2b \exp\{i[ cX-(c^2-b^2)T -\varphi]\} \times \\
\frac{(2b^2T-i) \cosh[b(X-2cT-X_0)] +ib(X-2cT) \sinh[b(X-2cT-X_0)]
}{\cosh^2[b(X-2cT-X_0)] + b^2[(X-2cT)^2+4b^2T]},
\end{array}
\label{tb}
\end{equation}
where $b, c$, $\varphi$, $X_0$ are arbitrary constants.

When the interatomic interaction is repulsive, i.e., $a_s(t)> 0$,
there are dark $N$-soliton solutions to Eq. (\ref{GP}). For the two
dark solitons solution, $\psi(x,t)$ is again expressed by Eq.
(\ref{SO}), but with $\phi(X,T)$ given by
\begin{eqnarray}
&\phi(X,T)= \tau_1 \exp\{i [l_1X -(l^2_1-2|\tau_1|^2)T
+\delta_1]\} \times \nonumber\\
&\frac{1 +\epsilon[Z_1 \exp(\xi_1) +Z_2 \exp(\xi_2)]
+\epsilon^2A_{12} Z_1 Z_2 \exp(\xi_1+\xi_2)}{ 1
+\epsilon[\exp(\xi_1) +\exp(\xi_2)] +\epsilon^2A_{12}
\exp(\xi_1+\xi_2)}, \label{td}
\end{eqnarray}
where $Z_1= ( -\Omega_1 +i\sqrt{4|\tau_1|^2 -\Omega^2_1} ) / (
\Omega_1 +i\sqrt{4|\tau_1|^2 -\Omega^2_1} )$, $Z_2= (-\Omega_2
+i\sqrt{4|\tau_1|^2-\Omega^2_2}) / (\Omega_2
+i\sqrt{4|\tau_1|^2-\Omega^2_2})$, $A_{12}= [(\Omega_1
-\Omega_2)^2 +(\sqrt{4|\tau_1|^2-\Omega_1^2} -\sqrt{4|\tau_1|^2
-\Omega_2^2})^2 ] / [(\Omega_1+\Omega_2)^2
+(\sqrt{4|\tau_1|^2-\Omega_1^2} -\sqrt{4|\tau_1|^2-\Omega_2^2})^2
]$, $\xi_1= (\sqrt{4|\tau_1|^2 -\Omega^2_1} -2l_1) \Omega_1 T
+\Omega_1X +\xi^{(0)}_1$, $\xi_2= (\sqrt{4|\tau_1|^2 -\Omega^2_2}
-2l_1) \Omega_2T +\Omega_2X +\xi^{(0)}_2$, $l_1$, $\delta_1$,
$\Omega_1$, $\Omega_2$, $\xi^{(0)}_1$, $\xi^{(0)}_2$, $\epsilon$
are real constants, and $\tau_1$ is complex constants.

\section{III. effects of the time-dependent magnetic fields on the solitons}

Now we consider the elementary applications of solutions
(\ref{SO}) with (\ref{tb}) and (\ref{td}) respectively, with
linear, exponential and sinusoidal time dependence of the magnetic
field via Feshbach resonance, and propose how to control dynamics
of solitons in BECs by synchronizing Feshbach resonance and
harmonic trap in future experiments.

\section{A. Magnetic field ramped linearly with time}
In real experiments \cite{Volz03,C. A. Regal}, the magnetic field
is linearly ramped down with time $t$. We can design an
experimental protocol to control the soliton interaction in BECs
near Feshbach resonance with the following steps: (i) In real
experiment of $^{87}$Rb atoms, the scattering length can be chosen
as a function of magnetic field, i.e., $a_s(t)= a_{\infty} [1-
\triangle/ (B(t)-B_0)]$ \cite{Volz03}, where the off-resonant
scattering length $a_{\infty}= 108 a_B$, $a_B$ is the Bohr radius,
$B_0$ is the Feshbach resonance position, and $\triangle$ is the
resonance width, respectively. The best-fit value for the width is
$\triangle=0.20$ G, resulting in $B_0=1007.40$ G. Near the
Feshbach resonance, the field $B(t)$ varies linearly with the rate
0.02 G/ms. For a better understanding, we plot Fig. 1, which shows
the scattering length and the trapping frequency vary with time
when the field $B(t)$ approaches the Feshbach resonance position
$B_{0}$. (ii) The realistic experimental parameters for a quasi-1D
repulsive condensate can be chosen $N \sim 10^3$ atoms and with
peak atomic density $n_0=10^5 cm^{-1}$. Then the scattering length
$a_s$ is of order of nanometer, e.g., $|a_s|=5.8$ nm for a
$^{87}$Rb condensate, and $\omega_{\bot}= 2\pi \times 400$ Hz,
with the ratio $\omega_1/ \omega_{\bot}$ being very close to zero.

The validity of the GP equation relies on the condition that the
system be dilute and weakly interacting: $n|a_s|^3 \ll 1$, where
$n$ is the average density of the condensate. Applying the above
conclusions to real experiments, we need to examine whether the
validity condition for the GP equation can be satisfied or not. In
the ground state for $^{87}$Rb condensate, the scattering length
is known to be $|a_s|=5.8$ nm \cite{Volz03}; the typical value of
the density ranges from $10^{13} - 10^{15}$ cm$^{-3}$. So
$n|a_s|^3< 10^{-3} \ll 1$ is satisfied. Moreover, the experimental
data agree reasonably well with the mean-field results \cite{D. S.
Jin}, which further proves the validity of the GP equation with
$|a_s|=5.8$nm. Another important issue is quantum depletion of the
condensate, which is ignored in the derivation of the GP equation.
The physics beyond the GP equation should also be very rich, and
we will work on more rigorous solutions beyond the GP equation in
the future.
\begin{figure}
\centering
\includegraphics[width=4.27cm,height=4.0cm,clip]{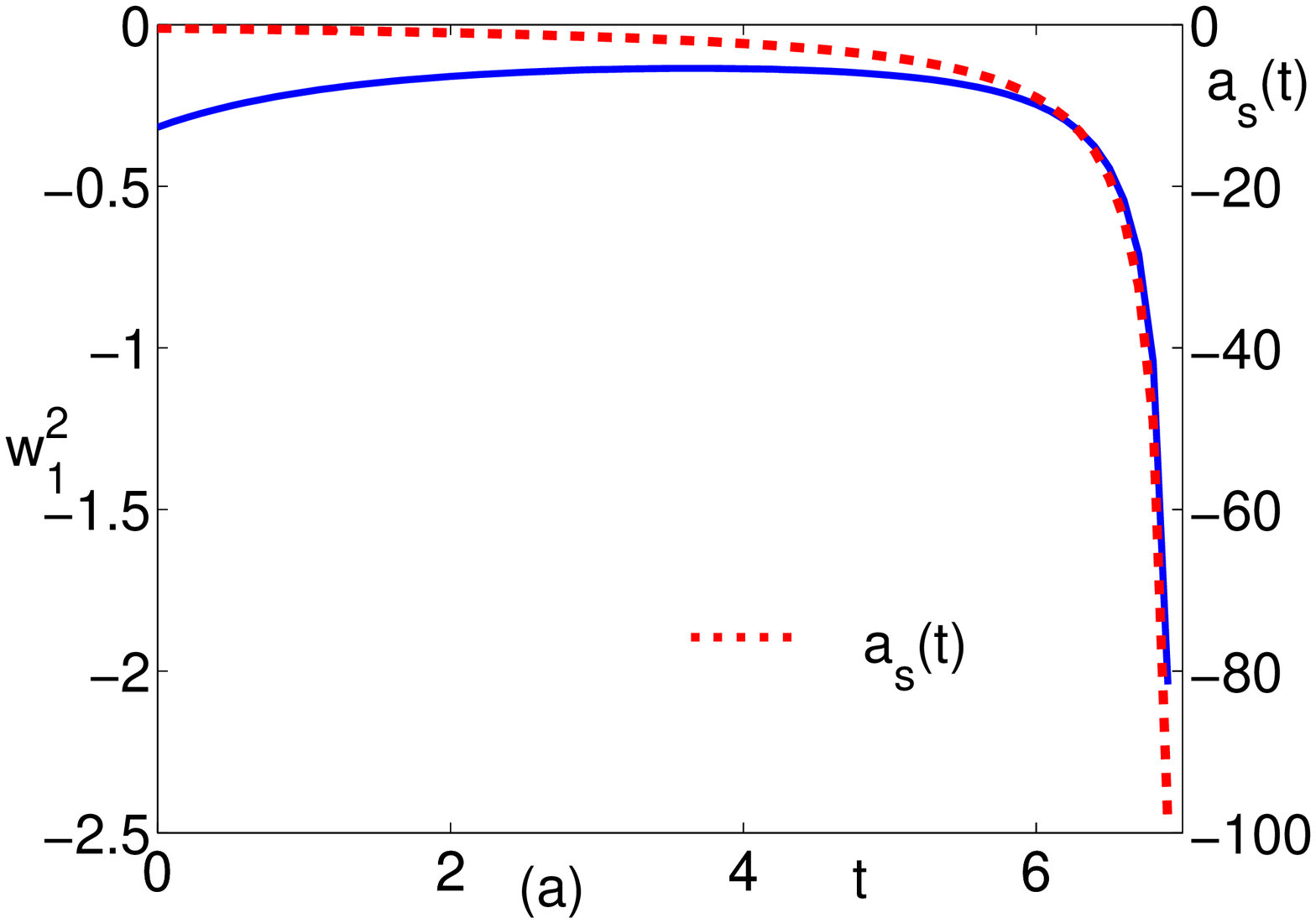}%
\hspace{0.1cm}%
\includegraphics[width=4.27cm,height=4.0cm,clip]{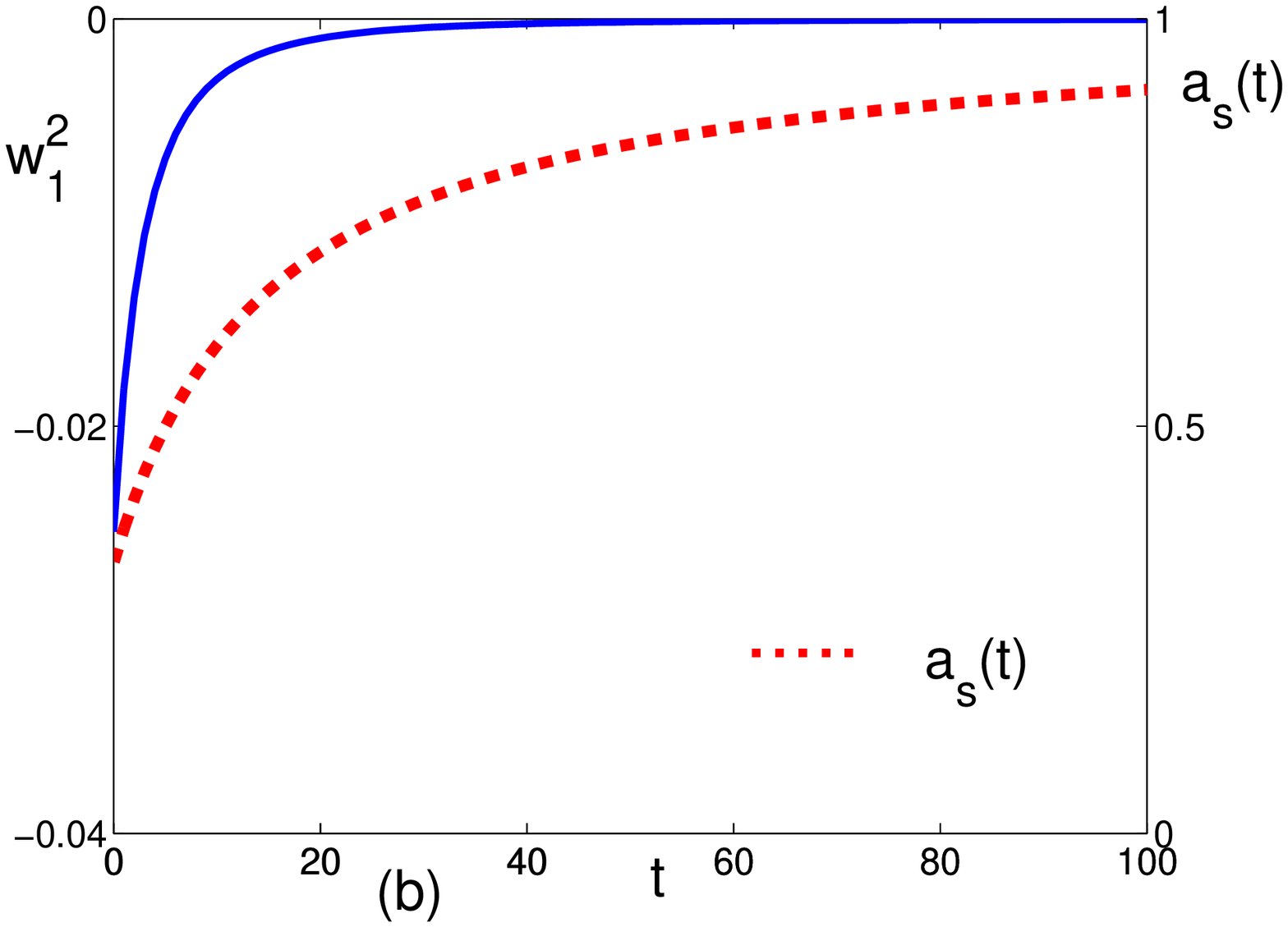}%
\hspace{0.1cm}%
\caption{(color online) The scattering length (red dotted line)
and the trapping potential (blue solid line) vary with time when
the magnetic field ramps linearly with time $t$. (a) The selected
field $B(t)$ varies from 1007.54 G to 1007.42 G, corresponding to
the attractive interaction between atoms. (b) The selected field
$B(t)$ varies from 1007.70 G to 1008.50 G, corresponding to the
repulsive interaction between atoms. The scattering length is
measured in units of $a_{\infty}$, the axial frequency is measured
in units of $w_{\bot}^2$, and the field $B(t)$ varies linearly with
the rate 0.02 G/ms.}
\end{figure}

Figure 2(a) shows the bright solitons interaction in BECs near the
Feshbach resonance. In this case, the $s$-wave scattering length
$a_{s}<0$. According to the integrability relation, the
time-dependent axial frequency is imaginary which indicates a
repulsive potential. As shown in Fig. 1(a), when the field $B(t)$
approaches the Feshbach resonance position $B_0$, the absolute
value of the scattering length increases, but the time-dependent
axial frequency decrease linearly. With the increasing of the
absolute of the scattering length, the interactions between atoms
become stronger, the peak of each soliton increases and its width
decreases. Meanwhile, under the expulsive potential, the two
bright solitons will be set into motion. As a result, the left
bright soliton feels two forces which come from the right bright
soliton and the trapping potential, it drives the left bright
soliton to the right-hand side. The right-hand one moves slower
than the left-hand one. Finally, the distance between the solitons
becomes smaller. When the field $B(t)$ infinitely approaches $B_0$, the absolute
value of the scattering length and axial frequency become
infinite. The two solitons interact very strongly and almost merge
into one with a very high peak and the narrowest width. After at
least close to 1007.50 G, the absolute value of the atomic
scattering length becomes $|a_s(t)|=5.7$ nm $<5.8$ nm for quasi-1D
$^{87}$Rb gas mentioned above. This means that the stability of
soliton and the validity of 1D approximation is maintained from
1007.54 G to 1007.50 G. With further increasing of $|a_s|$ [for
example, while $B(t)=1007.46$G, $|a_s|$ should be $13.3$nm], the
system may be beyond the validity of the GP equation. Therefore, the
phenomena discussed in Fig. 2(a) should be observable within the
current experimental condition from 1007.54 G to 1007.50 G. With
synchronized Feshbach resonance and harmonic trap to change the
scattering length and axial frequency, we can easily control
matter wave soliton interactions and obtain a new type of atom
laser with manipulatable intensity.

\begin{figure}
\centering
\includegraphics[width=6.0cm,height=5.0cm,clip]{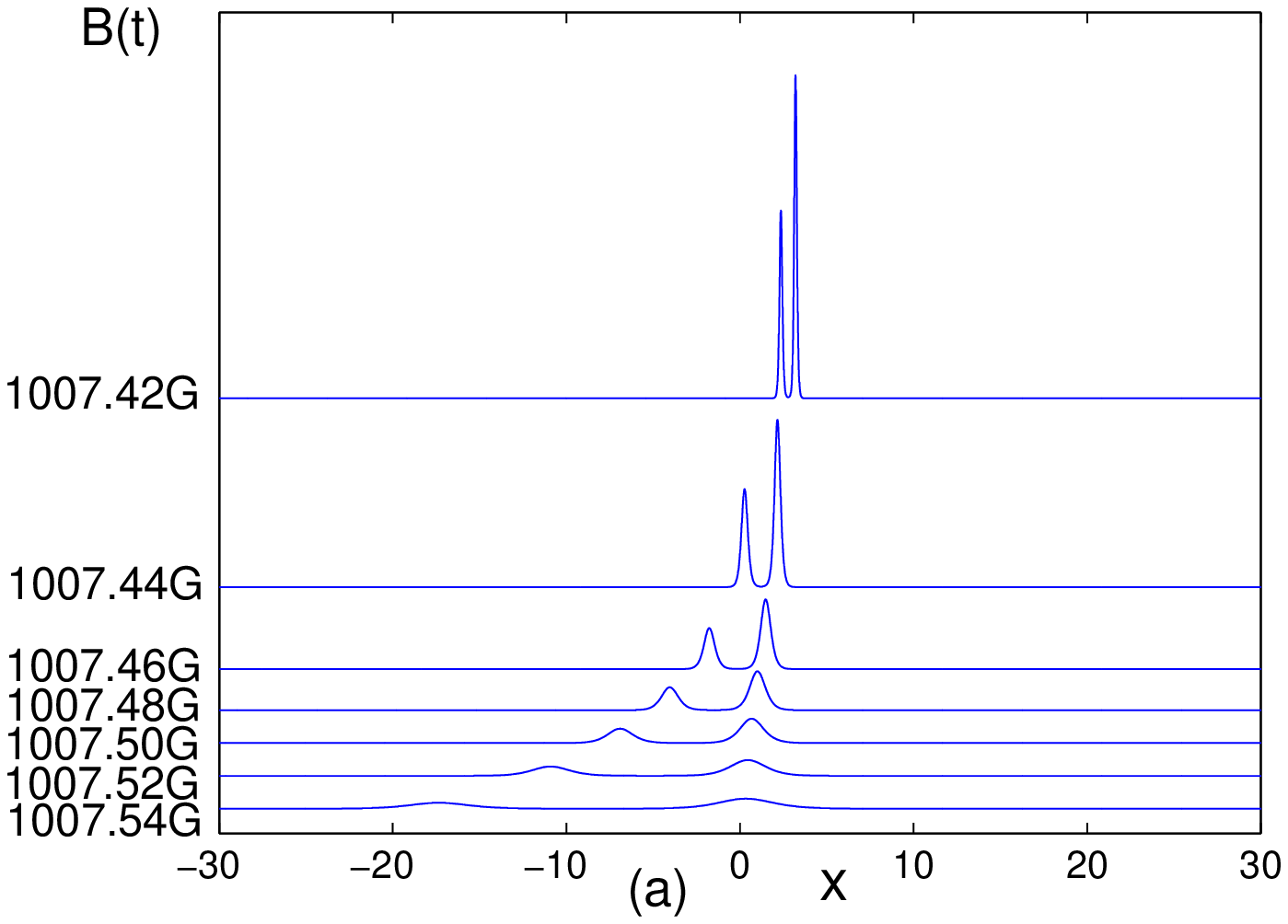}%
\hspace{0.1cm}%
\includegraphics[width=6.0cm,height=5.0cm,clip]{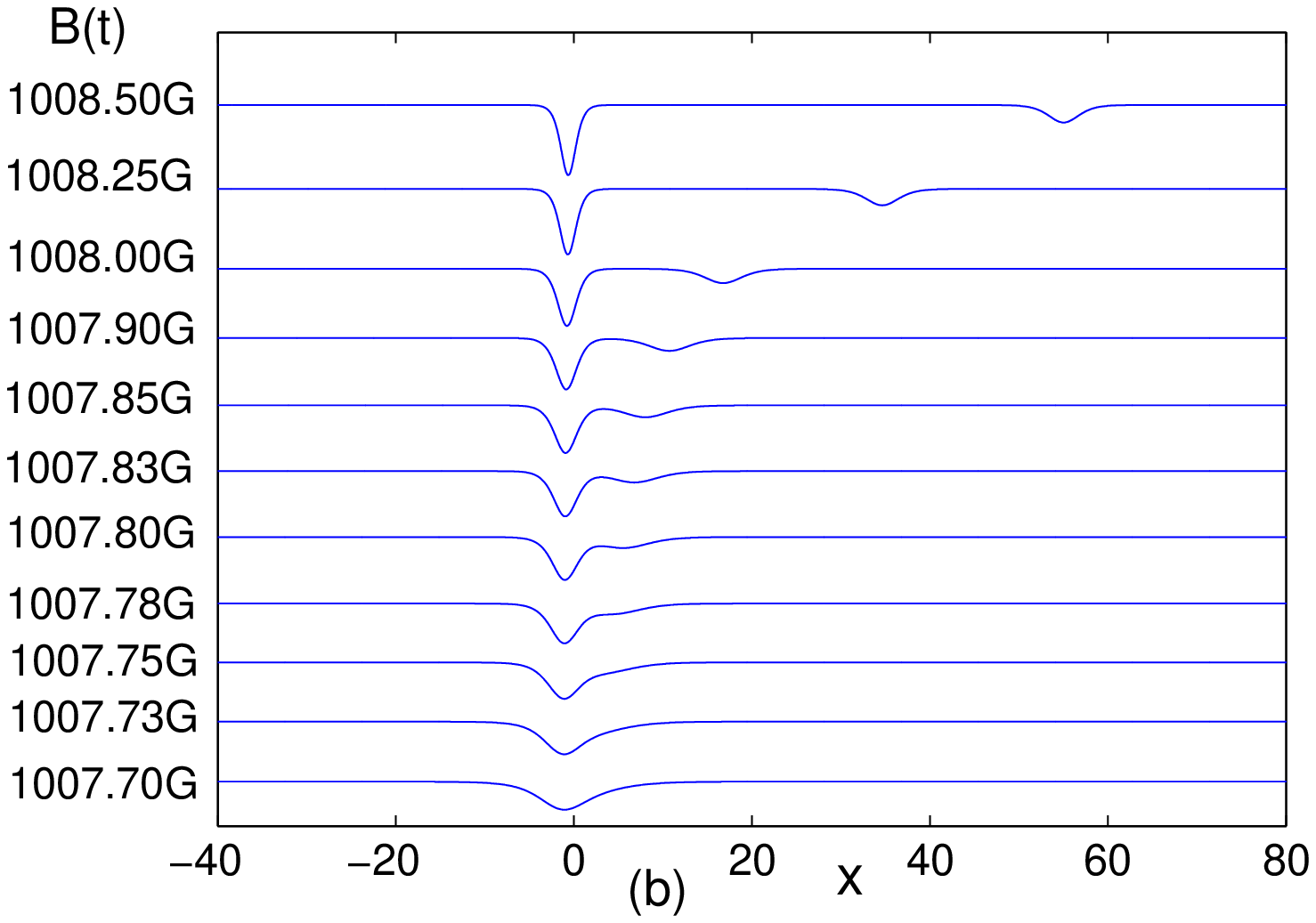}%
\hspace{0.1cm}%
\caption{(color online) Controlling matter wave bright and dark
soliton interaction near the Feshbach resonance ($B_0 =1007.40$ G)
when the magnetic field ramps linearly with time $t$. (a) The
selected field $B(t)$ varies from 1007.54 G to 1007.42 G. The
parameters are given as follows: $b=0.01$, $X_0=3000$, $c=0.002$
in Eq. (\ref{SO}) with Eq. (\ref{tb}). (b) The selected field
$B(t)$ varies from 1007.70 G to 1008.50 G. The parameters are
given as follows: $|\tau_1|^2=0.0001$, $\Omega_1=0.01$,
$\Omega_2=-0.02$, $\xi^{(0)}_1=112$, $\xi^{(0)}_2=0$, $\epsilon=1$
in Eq. (\ref{SO}) with Eq. (\ref{td}). }
\end{figure}

Figure 2(b) shows the interaction of dark solitons with magnetic
field being selected in the range from 1007.70 G to 1008.50 G. In
this case, as shown in Fig. 1(b), the $s$-wave scattering length
$a_{s}>0$, and is in proportion to the magnetic field, the axial
time-dependent decrease linearly and asymptotic approaches to
zero, it means that the system becomes a self-confined condensate.
Initially, there is only one dark soliton in BECs. Repulsive
interaction between atoms becomes stronger when the absolute value
of $a_{s}$ increases. This causes the dark soliton to split and
finally become two dark solitons, meanwhile, each soliton increases
its peak and compresses its width. However, contrary to the bright
soliton shown in Fig. 2(a), one dark soliton moves away from the
other. After at least close to 1008.50 G, the absolute value of
the atomic scattering length becomes $|a_s(t)|=4.67$ nm, which is
safely smaller than 5.8 nm for quasi-1D $^{87}$Rb condensate, thus
the stability of the soliton and validity of the 1D approximation are
maintained. We conclude that the dark soliton fission phenomenon
revealed here can be realized under the current experimental
condition.

\section{B. magnetic field varying exponentially with time}
When the field is exponentially ramped down as $\exp(-t/\tau)$ to
a selected field between 545 G and 630 G \cite{K. E. Strecker},
where $\tau=40$ ms, $a_s$ small and negative or small and
positive, the interaction parameter $g(t)$ near the resonance
varies exponentially with time: $g(t) \equiv a_s/a_B = \pm 0.01
\exp(\lambda t)$, where $\lambda = |\omega_1|/ \omega_{\bot}\ll1$.
The integrability relation reads
$-\lambda^{2}=\frac{w_{1}^{2}}{w_{\bot}{2}}$, and can be satisfied
automatically as long as the time-dependent axial frequency is
imaginary which indicates a repulsive trapping potential. With the
same parameters as in the experiment \cite{L. Khaykovich}, i.e., $N
\approx 10^3$, $\omega_{\bot}= 2\pi \times 700$ Hz, $\omega_1=
2i\pi \times 21$ Hz, and $\lambda= 0.03$, the interactions between
the two bright solitons is shown in Fig. 3. It is interesting to
observe that in the expulsive parabolic potential, the bright
solitons are set into motion and propagate in the axial direction.
With time going on, $|a_s(t)|$ increases. We could observe an
increase in their peaking values and a compression in widths,
besides that the spacing between them decreases after they are
initially generated on different positions in the trap, which is
evidence for a short-range attractive interaction between
solitons. Finally, they almost merge and fusion. This phenomena is
different from the case of \cite{K. E. Strecker, L. Khaykovich}.
There, the bright solitons are set in motion by off setting the
optical potential and propagate in the potential for many
oscillatory cycles with the period 310 ms, the spacing between the
solitons increase near the center of oscillation and bunches at
the end points. The difference is mainly caused by two factors:
one is the time-varying scattering length strongly affect the
interaction between the solitons, and the other reason is the
repulsive force provided by the potential. Meanwhile, with the
increasing of the scattering length, the attractive interaction
between atoms become stronger, this will leads to an attractive
interaction between solitons. Eventually, the outcome of these two
forces will determine the motion of the two bright solitons.

Fusion is very interesting phenomenon and it comes from the
interatomic attractive interaction. In other words, with time
going on, both bright solitons change their positions, warp in a
certain radian, and almost merge into one single soliton. Such
morphology has been observed in coronal plasma \cite{L. Golub}. We
hope that such morphology would be detected in BEC experiments too
in the near future.

\begin{figure}[tbp]
\centering
\includegraphics[width=6.00cm]{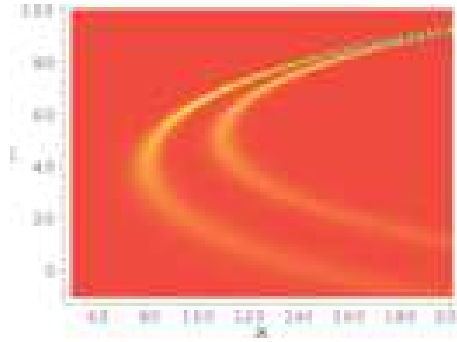}\hspace{1.0cm}
\caption{(color online) Controlling matter wave bright soliton
interaction when field varying exponentially with time. The
parameters are given as follows: $\lambda=0.03$, $b=0.5$, $X_0=20$,
$c=7$ in Eq. (\ref{SO}) with Eq. (\ref{tb}). The color corresponds
to the BEC density, with yellow (gray) being the smallest and blue
(dark) being the largest.}
\end{figure}

In real experiments \cite{K. E. Strecker}, the length of the
background of BECs can reach at least $2L= 370$ $\mu$m. At the
same time, in Fig. 3, solitons travel from $x=$ 50 to 200, i.e.,
$150 \times 1.4$ $\mu$m$= 210$ $\mu$m. [The dimensionless unit of
the coordinate, $\Delta x= 1$, corresponds to $a_\bot= (\hbar/m
\omega_\bot)^{1/2} = 1.4$ $\mu$m]. We indeed have $210$ $\mu$m $<
370$ $\mu$m, a necessary condition for observing the morphology in
BEC experiments \cite{Z. X. Liang}. Additionally, after at least
up to 100 dimensionless units of time, $|a_s(t)|$ reaches the
value $0.2a_B$, which is less than $|a_{final}| = 4a_B$. This
means that during the time evolution, the stability of solitons
and the validity of 1D approximation can be maintained as
displayed in Fig. 3. Therefore, the phenomena discussed in this
case are also expected to be observable within the current
experimental capability.

The interactions between two dark solitons are also intriguing.
The first experimental evidence of attraction between dark
solitons in nonlocal nonlinear media has been presented \cite{A.
Dreischuh}. Our results (3), (5), and (7) also indicated that
attraction between dark solitons should be observable in BECs with
repulsive long-range interatomic interaction. In a previous case,
the field is linearly ramped down as time, leading to the trap
axial frequency $\omega_1 \cong 0$ and external potential
vanishing. It means that the system becomes a self-confined
condensate. So the repulsion between the dark solitons is observed
in BEC with repulsive interatomic interaction [seen in Fig. 2(b)].
However, in the present case, the field varies exponentially with
time, the trap axial potential is not vanishing and
time independent, the system is in an expulsive parabolic
potential. Following the experimental setup in \cite{L.
Khaykovich}, we can first create a BEC in the quasi-one-dimensional
potential. Second, the trap potential is tuned to the value in our
paper, meanwhile, the scattering length varies exponentially with
time to a small and positive value. The potential provides an
attractive force which counters the natural repulsion of the
solitons. Finally, the competition between these two forces will
determine the outcome of the interaction between solitons. If the
attractive force which is caused by the potential is stronger than
the natural repulsion of the solitons, the two dark solitons will
move towards the other, which is the evidence for a short-range
attractive interaction between dark solitons. However, the
lifetime of a BEC in current experiments is of the order of 1 s
and the region of a BEC is small, the solitons will be dissipation
in their motion before reaching the edge of the condensate.
\begin{figure}[tbp]
\centering
\includegraphics[width=6cm]{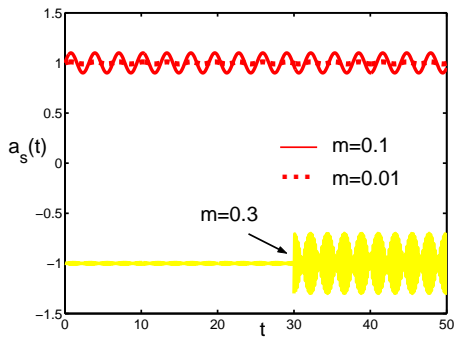}\hspace{1.0cm}
\caption{(color online) The scattering length varies with time
when field varying periodically with time $t$. The yellow curve
(gray) denotes the attractive interaction case, at $t=30$ (marked
by an arrow), the amplitude of the scattering length is larger than
before by suddenly jumping the amplitude of the ac drive $m$ to
$m=0.3$, $w=30$. The red curve denotes the repulsive interaction
for $m=0.1$ (dark solid line) and $m=0.01$ (dark dotted line),
$w=2$. All of the lengths are measured in units of $a_{B}$.}
\end{figure}
\section{C. magnetic field varying periodically with time}
It was observed that a small sinusoidal modulation of the magnetic
field close to the Feshbach resonance gave rise to a modulation of
the interaction strength, $g(t) \equiv a_s(t)/a_B = \pm [1 +
m\sin(\omega t)]$ \cite{M. Greiner}, where the amplitude $m$ of
the ac drive was small and satisfied $0< m <1$. According to the
integrability condition, the axial frequency of the harmonic
potential should be
\begin{equation}
\omega^2_1(t)= -\frac{m\omega^2 \omega^2_{\bot}}{[1+ m\sin(\omega
t)]^2} [\sin(\omega t) +m + m\cos^2(\omega t)].
\end{equation}

Now, we investigate how the amplitude of the ac drive can be used
to control the bright soliton interactions. As shown in Fig. 6,
for the case with atomic attractive interaction, when the
amplitude is small, $m= 0.01$ ($0< t <30$). The periodically
varying of the scattering length and the trapping frequency are
small as shown in Figs. 4 and 5, the repulsive and attractive
force between solitons can become balanced. Two bright solitons
will move in parallel with their separation keeping constant. This
property would be interesting for optical communication with low
bit-error rates. In BEC, this effect may play an important role in
potential application of matter wave communication with atom
lasers. If the amplitude is increased to the value $m=0.3$
($30< t <50$), the solitons begin to oscillate due to the temporal
periodic modulation of the $s$-wave scattering and trapping
potential are both stronger than before. This phenomena is very
similar to the evolution of optical solitons.
\begin{figure}
\centering
\includegraphics[width=3.9cm,height=4.2cm,clip]{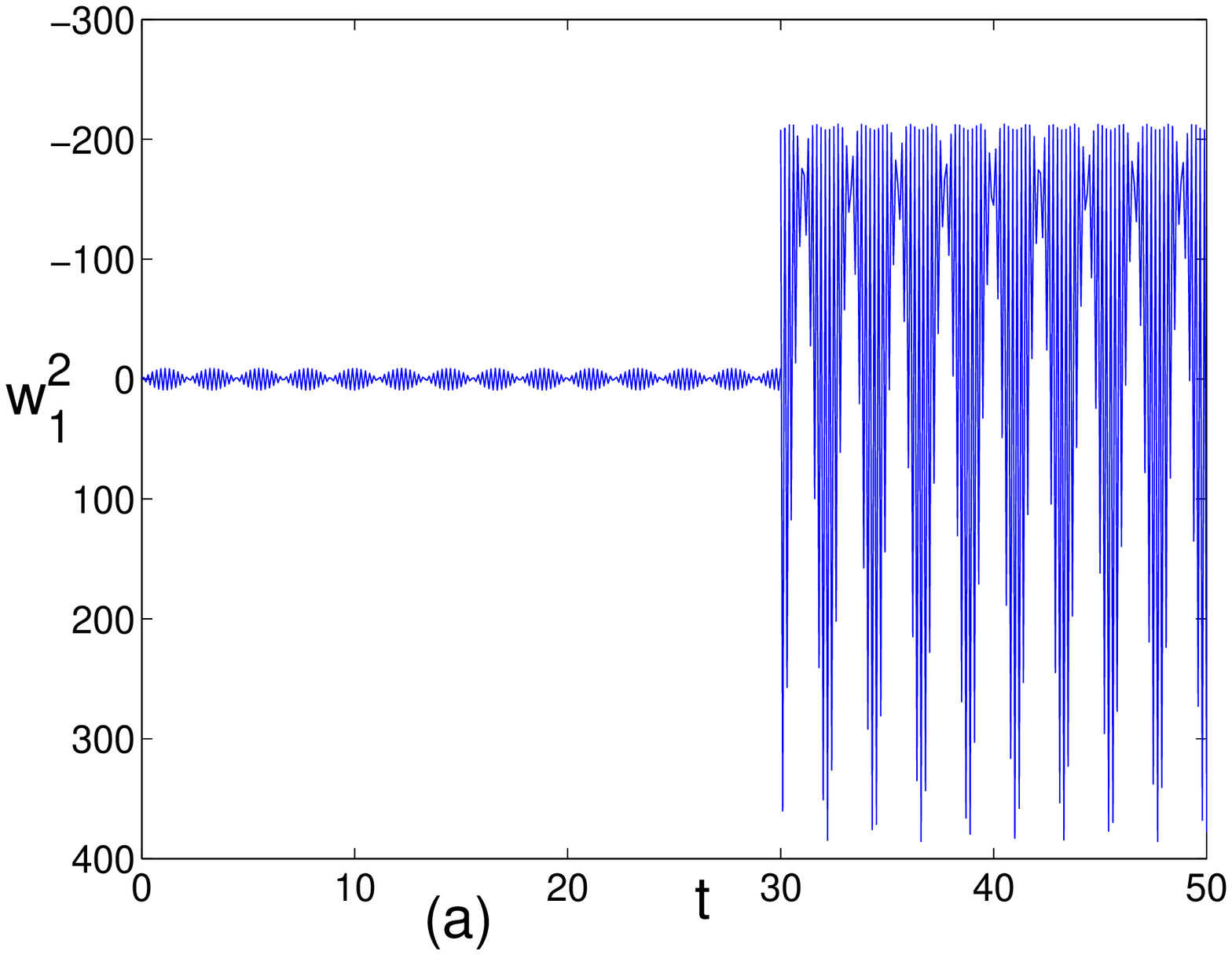}%
\hspace{0.2cm}%
\includegraphics[width=4.2cm,height=4.2cm,clip]{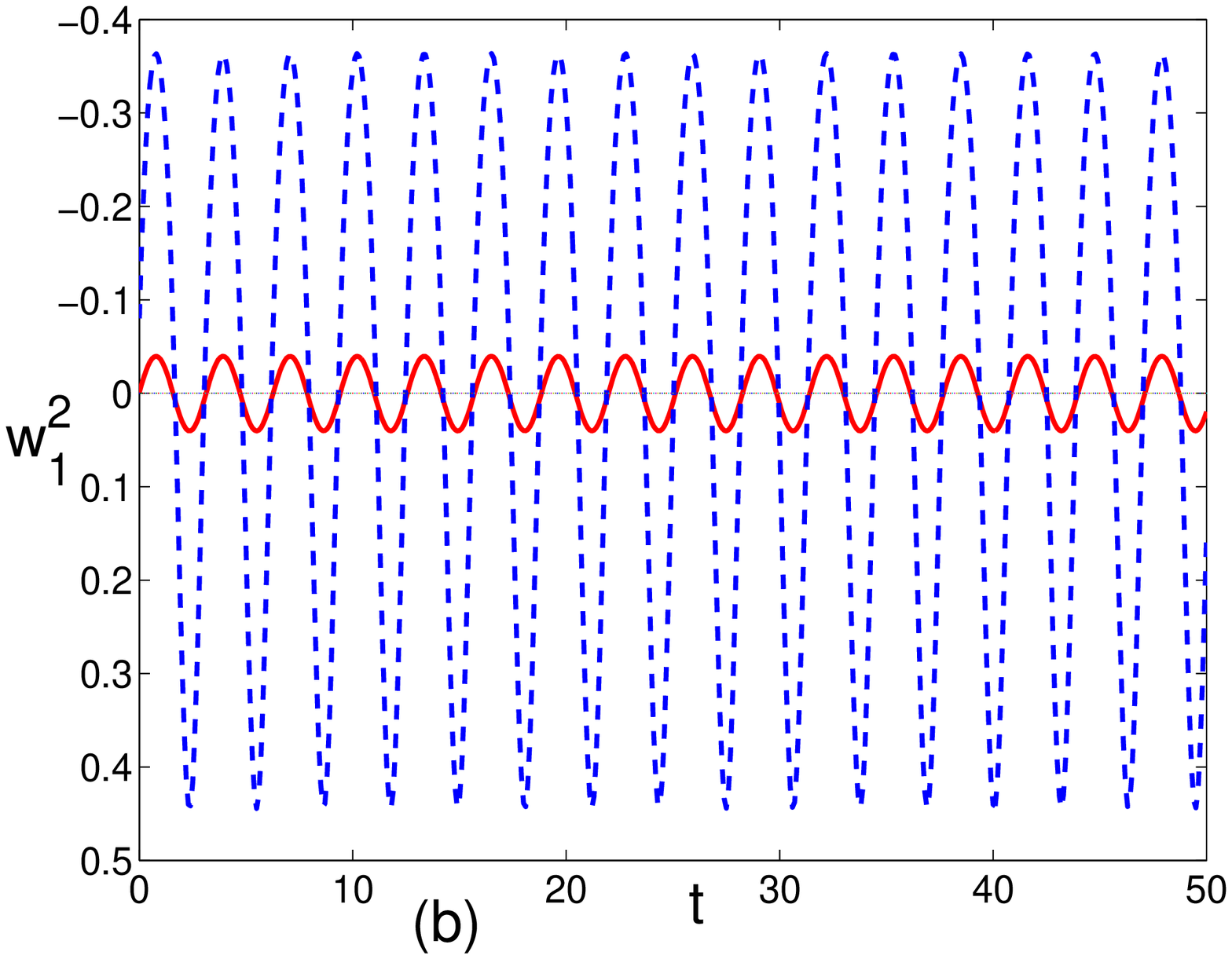}%
\hspace{0.2cm}%
\caption{(color online) The trapping frequency varies with time
when field varying periodically with time $t$. (a) $m=0.01$
($0<t<30$), $m=0.3$ ($30<t<50$), and $w=30$ for the case with
atomic attractive interaction. (b) $m=0.01$ (red solid line),
$m=0.1$ (blue dashed line), and $w=2$ for the case with atomic
repulsive interaction. The frequency is measured in units of
$w_{\bot}^2$.}
\end{figure}

We also studied the effect of ac drive on interactions between the
dark solitons. First, the amplitude $m$ of the ac drive is
chosen small, $m = 0.01$, this leads to a small periodical
modulation of the scattering length and the trapping frequency.
The repulsive interaction between atoms mainly leads to the
formation of the dark solitons, this force cannot lead to a
oscillation of the dark solitons due to the small change in the
scattering length. As shown in Fig. 7(a), a faster soliton is
generated behind a slower one in the mutual moving direction,
after an interval of time, the faster one pulls up to the slower
one and their elastic collision happens. After the collision, the
faster one passes through the slower one and their parameters did
not change, which remarkably indicates no energy exchange between
the two dark solitons. For larger ac drive, for example, $m =
0.1$, which is 10 times of the value of the previous $m$, with the
same initial condition, because of the stronger modulation of the
scattering length and the trapping frequency, both solitons move
forward, meanwhile, they will oscillate back and forth, as shown
in Fig. 7(b). In all of the above cases, the trap potential can become
repulsive during the entire process, but the attractive potential
is stronger than the repulsive one during a period, the outcome of
the potential is an attractive one which can be seen in Fig. 5. It
shows that a BEC with repulsive interaction between atoms is
confined in the trap, this is different from the case B.

Another important problem is the atom loss. As the external
magnetic field is driven close to the resonant value, the rate of
loss of atoms is increasing rapidly in the vicinity of the
Feshbach resonance, while only a small fraction of atoms remain as
soliton. In all of the above cases, the scattering length is small, the
validity of the GP equation is satisfied. Meanwhile, when the
trapping potential is modulated according to the integrability
relation, both the rate of the untrapped atom and the collective
excitations will be further discussed.
\begin{figure}[tbp]
\centering
\includegraphics[width=6cm]{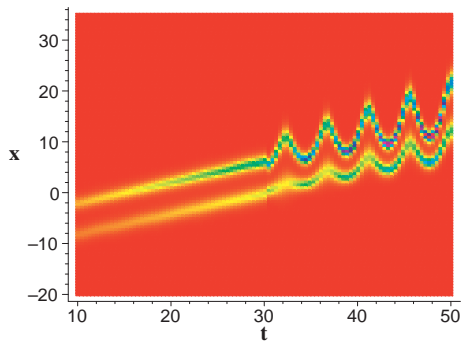}\hspace{1.0cm}
\caption{(color online) Controlling matter wave bright soliton
interaction when field varying periodically with time, where (a)
$m=0.01$ ($0<t<30$), (b) $m=0.3$ ($30<t<50$). The other parameters
are as follows: $\omega=30$, $b=1$, $X_0=-9$, $c=0.4$ in Eq.
(\ref{SO}) with Eq. (\ref{tb}).}
\end{figure}
\begin{figure}
\centering
\includegraphics[width=6.0cm,height=5.0cm,clip]{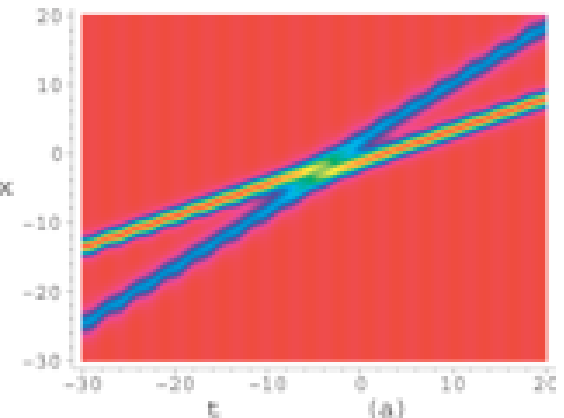}%
\hspace{0.1cm}%
\includegraphics[width=6.0cm,height=5.0cm,clip]{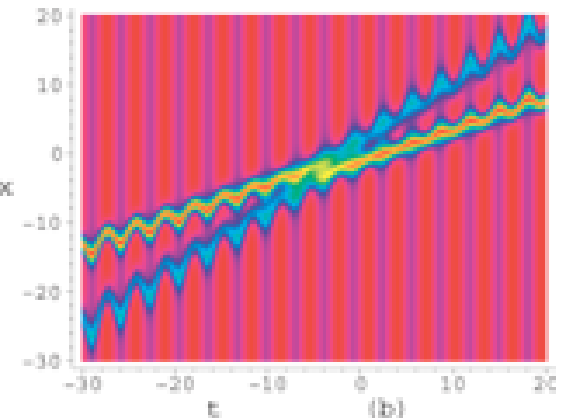}%
\hspace{0.1cm}%
\caption{(color online) Controlling matter wave dark soliton
interaction when field varying periodically with time, where (a)
$m=0.01$, (b) $m=0.1$. The other parameters are as follows:
$\omega=2$, $|\tau_1|^2=1$, $\Omega_1=1.2$, $\Omega_2=-1.8$,
$l_1=0.02$, $\xi^{(0)}_1=-0.01$, $\xi^{(0)}_2=-0.2$, $\epsilon=1$
in Eq. (\ref{SO}) with Eq. (\ref{td}).The red (with smaller slope)
and blue correspond to the slower one and the faster one
respectively.}
\end{figure}

\section{IV. conclusion}
In summary, we express how to control soliton interaction in BECs
with arbitrary time-varying scattering length in a synchronized
time-dependent harmonic trap. When the integrability condition is
satisfied, we obtained the exact solutions analytically, and
explored the interaction of the bright and dark solitons in BECs
with Feshbach resonance magnetic field linearly, exponentially, and
periodically dependent on time. In these typical examples, we find
several interesting phenomena involving soliton interactions, such
as fusion, fission, warp, oscillation, elastic collision, etc.
\cite{F. K. Abdullaev, G. Theocharis}. We further discussed how to
control interactions between bright or dark solitons, in BECs in
realistic situations, which allows for experimental test of our
predictions in the future. These phenomena open possibilities for
future applications in coherent atom optics, atom interferometry,
and atom transport.

This work was supported by NSFC Grants Nos. 90406017, 60525417 and 10740420252 and NKBRSFC Grants Nos.
2005CB724508 and 2006CB921400.

\end{document}